# Experimental Realization of Acoustic Chern Insulator

Yifan Zhu[1†], Yugui Peng[2†], Xudong Fan[1], Jing Yang[1], Bin Liang[1*], Xuefeng Zhu[2*]

and Jianchun Cheng[1*]

[1]*Key Laboratory of Modern Acoustics, MOE, Institute of Acoustics, Department of Physics, Collaborative Innovation Center of Advanced Microstructures, Nanjing University, Nanjing* 210093, *P. R. China*

[2]*School of Physics and Innovation Institute, Huazhong University of Science and Technology, Wuhan, Hubei* 430074, *P. R. China*

[†]These authors contributed equally to this work.

[*]To whom correspondence should be addressed. Emails: liangbin@nju.edu.cn (B.L.), xfzhu@hust.edu.cn (X.F.Z.), jccheng@nju.edu.cn (J.C.C.)

**Abstract**

Topological insulators are new states of matter in which the topological phase originates from symmetry breaking. Recently, time-reversal invariant topological insulators were demonstrated for classical wave systems, such as acoustic systems, but limited by inter-pseudo-spin or inter-valley backscattering. This challenge can be effectively overcome via breaking the time-reversal symmetry. Here, we report the first experimental realization of acoustic topological insulators with nonzero Chern numbers, *viz.*, acoustic Chern insulator (ACI), by introducing an angular-momentum-biased resonator array with broken Lorentz reciprocity. High Q-factor resonance is leveraged to reduce the required speed of rotation. Experimental results show that the ACI featured with a stable and uniform metafluid flow bias supports one-way nonreciprocal transport of sound at the boundaries, which is topologically immune to the defect-induced scatterings. Our work opens up opportunities for exploring unique observable topological phases and developing practical nonreciprocal devices in acoustics.

Symmetries play a fundamental role in many physical phenomena we observe. For example, quantum spin Hall effect preserves time-reversal symmetry, for which the subsequent topological explanation gives birth to one of the breakthroughs in condensed matter physics, *viz.*, topological insulators [1-5]. Recently, the time-reversal symmetry protected topological insulators have been readily extended to classical wave systems [6-16], such as acoustic systems [10-16]. To trigger topological phase transition in acoustic systems, spatial symmetry breaking is employed to gap the Dirac degeneracy in momentum space and give rise to the acoustic analog of spin or valley degree of freedom [10-13]. It is well known that the static and linear acoustic systems obey bosonic-like time-reversal symmetry $T_b^2 = 1$. Unlike spin-1/2 particles with fermionic time-reversal symmetry $T_f^2 = -1$, the bosonic-like nature of sound indicates that wave components carrying different pseudo-spins or valleys can be converted into one another via the time-reversed channels, thus leading to the problem that inter-pseudo-spin or inter-valley scattering is inevitable. To overcome this issue, it is intuitive to close all time-reversed channels by breaking the Lorentz reciprocity [17-22]. In previous works, nonreciprocal sound propagation was implemented through nonlinearity or magneto-acoustic effect [23-24], but only with the drawbacks of low efficiency and large volume. In contrast, giant acoustic nonreciprocity can be provided by an angular-momentum-biased resonant circulator. However, in the construction of nonreciprical topological insulator with uniformly biased ciruclators, other serious challenges emerge, such as nonsynchronous rotation and flow instabilities, making its practical implementation hitherto elusive.

In this Letter, we propose to use a rotating chiral structure to generate a stable and uniform metafluid flow that is circulating unidirectionally in a three-port ring resonator. Combining high Q resonances with specially designed ports, this elegant approach largely reduces the required speed of rotation for producing giant acoustic nonreciprocity, and meanwhile ensures the uniformity of angular momentum bias, the homogeneity of flow field, and the controllability of flow velocity in each resonator, in stark contrast with previous methods [24]. On the basis of this novel configuration, we experimentally demonstrate the prototype of acoustic Chern insulator (ACI), *viz.*, a 'phononic graphene' comprising coupled ring resonators with uniformly biased flow field. The proposed model presents an acoustic analog of integer quantum Hall effect, where the Chern number is nonzero due to the existence of effective magnetic field (see Supplementary Note 1).

Figure 1(a) shows the schematic diagram of ACI. The ACI is a honeycomb lattice of ring resonators (artificial 'atoms') connected by waveguides. A primitive cell is shown by a shaded rhombus region, which contains two artificial 'atoms' A and B. The top view of a unit cell is displayed in Fig. 1(b), which comprises two resonators with different orientations of the connection ports. Inner cores of ring resonators with chiral-structured fins are rotors driven by motors of tunable rotational speed. The motor supports both clockwise and anticlockwise rotations, defined as positive and negative revolutions, respectively. Figure 1(c) shows the photograph of ring resonator sample, where the chiral-structured fins of rotational symmetry are clearly presented. For structrual parameters of the ring resonators, height, inner and outer radii of the rotor are $H$=4 cm, $R_{out}$=9.2 cm and $R_{in}$=7.1 cm. The waveguide width $W$=5 cm and the distance between two neighboring resonators (*e.g.*, A and B) $P$=40 cm. From the inset of Fig. 1(b), the angle between the radial direction and the tilted fin $\vartheta$=49º. The width and length of the fin are $w$=0.3 cm and $l$=1.6 cm. Figures 1(d) and 1(e) show the three-port resonators with rotors fixed on motionless and rotating motors, respectively. The frequency of rotation is 5 rev/s. The inclined fins on the rotors can be effectively regarded as an annular metafluid for long waves, which, according to aerodynamics, helps to generate a stable air flow in rotation. The air flow simulation is provided in the Supplementary Note 2, Fig. S1, showing that the chiral structure is optimized to generate the desired and efficient flow field, and is a very good candidate to build up a uniform and controllable angular-momentum-biased array for ACI. The experiments are operated in an anechoic chamber, where the prepared samples are inserted between two parallel plexiglass plates and the motors are fixed under the bottom plate. Experimental details are appended in the Supplementary Note 3. Ring resonators can in principle support different orders of whispering gallery modes [25]. Here, we choose the higher-ordered mode (*e.g.*, octupole resonance in our case) that provides a higher Q factor of resonance and requires lower frequency of rotation for producing giant acoustic nonreciprocity than the lower-ordered resonance modes. Numerical simulations prove that for quadripole and octupole resonances, the maximum nonreciprocal acoustic propagations will occur at 20 rev/s and 5 rev/s, respectively (see Supplementary Note 2, Figs. S2(a-b)). In addition, it is proved that the variation of waveguide cross-section helps to reduce the required frequency of rotation. Through the numerical simulation of one biased resonator (see Supplementary Note 2, Fig. S2(e)), we obtain that when the cross-section raito $w_0/W$ changes from 0.1, 0.2, 0.3, to 0.4, the required frequencies of rotation at the maximum nonreciprocity are calculated to be ±5 rev/s, ±14.6 rev/s, ±25.5 rev/s, and ±38 rev/s, respectively, where $w_0$ refers to the width of connections between waveguides and ring resonators. Therefore, the local resonances introduced from the varied cross-section can largely decrease the required frequency of rotation, consistent with the fact that the non-resonant approach is impractial with an extremely high frequency of rotation (~200 rev/s) for the biased flow [17].

We start from studying the nonreciprocal sound propagation in a unit cell of ACI, which comprises two biased resonators at the frequency of rotation 5 rev/s. Figure 2(a) shows the photograph of one unit cell with four ports $P_a$, $P_b$, $P_c$, and $P_d$. In the experiment, acoustic waves are incident at port $P_a$, while the transmission spectra are extracted at ports $P_b$ and $P_c$. From Fig. 2(b), the measurement results show that asymmetric transmission at ports $P_b$ (purple circles) and $P_c$ (black circles) emerges within the range of 1930~1970 Hz. In this case, acoustic waves are inclined to output at port $P_b$, giving rise to higher sound transmission than that at port $P_c$. The maximum transmission contrast (~14) is observed at 1958 Hz, where octupole resonance takes place. In Figs. 2(c) and 2(d), the simulated and measured intensity distributions at 1958 Hz agree well with each other. At the resonance frequency, nearly all sound energy is transmitted to port $P_b$ instead of port $P_c$. This phenomenon is caused by the broken reciprocity in the angular-momentum-biased resonator. In Supplementary Note 4, Fig.

S4, we further present the simulated intensity distributions for acoustic waves incident at ports $P_a$, $P_b$, $P_c$, and $P_d$, respectively, where the nonreciprocal sound propagation is clearly demonstrated. In the following, we will show that non-reciprocity is the key to trigger topological phase transition in ACI.

Figure 3 unveils that the topological phase transition in ACI takes place as the rotors in resonators spin. The first Brillouin zone of the honeycomb lattice is illustrated in Fig. 3(a), where the irreducible Brillouin zone is marked by the triangle ΓMK. Based on finite element simulation (see Supplementary Note 3), we calculate the band structures of the honeycomb lattice with the frequency of rotation being 0 rev/s and 5 rev/s, respectively. The band diagram is plotted around the octupole resonance frequency 1958 Hz. Due to lattice symmetry and time-reversal symmetry, the band structure of unbiased honeycomb lattice is featured with the existence of Dirac cones at the K point, as shown by the black circles in Fig. 3(b). After applying a flow bias in the resonators to break time-reversal symmetry, the Dirac degeneracy at K point together with the two-fold degeneracies at Γ point are lifted, as shown by the purple circles in Fig. 3(b). The generated band gaps have nontrivial topological properties, which are associated with the change of Chern number of bulk bands. Here, the Chern numbers of four bulk bands (purple circles in Fig. 3(b)) are calculated to be −1, 0, 0, 1 from lower to upper bands (see Supplementary Note 3). When the rotational direction of rotors is reversed to anticlockwise, the Chern numbers of bulk bands change into 1, 0, 0, −1, corresponding to the case of reversing the constructed effective magnetic field.

To explore the property of topological edge mode, we present the band structure of a stripe super-cell comprising 10 biased resonators at the frequency of rotation 5 rev/s. From the band diagram in Fig. 3(c), we observe gapless edge modes (blue and red circles) between bulk modes (black circles). The gapless edge modes locate at 1926.7~1944.7 Hz and 1956.4~1965 Hz. In Fig. 3(d), we show the intensity field distributions of edge modes and bulk modes. Eigenfrequencies of edge modes and bulk modes are chosen at 1958 Hz and 1967 Hz, respectively. Intensity distributions reveal that the edge modes with positive group velocity (blue arrow) are localized at the bottom boundary, while the ones with negative group velocity (red arrow) are localized at the upper boundary. For bulk modes, the field is distributed across the whole lattice. It is worth to mention that a wider nontrivial band gap can be obtained if we increase the rotational speed of flow, which however enhances the difficulty in realizing uniformly biased resonator lattice. Considering the trade-off between lower rotational speed of rotor and wider observable nontrivial band gap, we choose 5 rev/s as the frequency of rotation in the experiment.

At last, we investigate the properties of ACI in experiments. Figure 4(a) shows the fabricated sample of ACI, comprising 4×7 resonators with the size of 2.5 m×3 m. In measurement, we extract the acoustic pressure amplitude at five probe ports P0, P1, P2, P3, and P4 on the honeycomb lattice as marked by the red dots in Fig. 4(a). The data at P1 and P2 are employed to distinguish edge modes and bulk modes, while at P3 and P4 measurements are conducted to further determine the chirality of edge modes. Here we define the measured pressure amplitudes at ports P1, P2, P3, and P4 to be $A_1$, $A_2$, $A_3$, and $A_4$, for which the spectra are shown in Figs. 4(b) and 4(c) with the frequencies of rotation being 5 rev/s and −5 rev/s, respectively. The frequency ranges of edge modes are covered by yellow ribbons, corresponding to the case of $A_1>A_2$, which locate at 1935.5~1944.5 Hz, 1955.0~1964.0 Hz in Fig. 4(b) (anticlockwise chiral edge modes for $A_3>A_4$) and 1933.0~1939.0 Hz, 1955.0~1967.5 Hz in Fig. 4(c) (clockwise chiral edge modes for $A_3<A_4$), respectively. These measured ranges agree with the numerical prediction in Fig. 3(c). From the experimental results, we find that nonreciprocal topological propagation of sound for clockwise rotation of rotors is as good as the one for anticlockwise rotation of rotors, which is demonstrated from the measurement of one positively or negatively biased resonator (see Supplementary Note 2, Figs. S2(c-d)). In Figs. 4(d-i), both simulated intensity fields and measured amplitude distributions unequivocally show the existence of anticlockwise chiral edge mode (1958 Hz, 5 rev/s), bulk mode (1967 Hz, 5 rev/s), clockwise chiral edge mode (1958 Hz, −5 rev/s). Here we point out that the acoustic loss is an inevitable issue in experiment, approximately 1 dB attenuation per a biased resonator. Therefore, in Figs. 4(b-c), loss compensation is added to the measured spectra at ports P3 and P4 for a better comparison among $A_1$, $A_2$, $A_3$, and $A_4$. It needs to be mentioned that other nonreciprocal propagation with topological protection can be expected, such as topological interface modes between two ACIs with reversed angular momentum biases (see Supplementary Note 5, Fig. S5).

In summary, we have experimentally demonstrated the first prototype of ACI, where the topological phase transition is induced by uniform angular momentum bias in the ring resonators. In the topologically nontrivial band gaps, we observe nonreciprocal chiral edge modes with strong immunity to backscattering at boundary defects. Our approach featured with low rotational speed of biased flow paves the way towards the production of nonreciprocal acoustic devices with topologically protected unidirectional transport, as required for various key applications in acoustic communications, noise control, and medical ultrasonics.

**Acknowledgements.**

This work was supported by the National Key R&D Program of China, (Grant No. 2017YFA0303700), National Natural Science Foundation of China (Grant Nos. 11634006, 81127901, 11674119, 11404125, and 11690030), and A Project Funded by the Priority Academic Program Development of Jiangsu Higher Education Institutions. X. F. Z and Y. G. P. acknowledge the financial support of Bird Nest Plan of HUST.

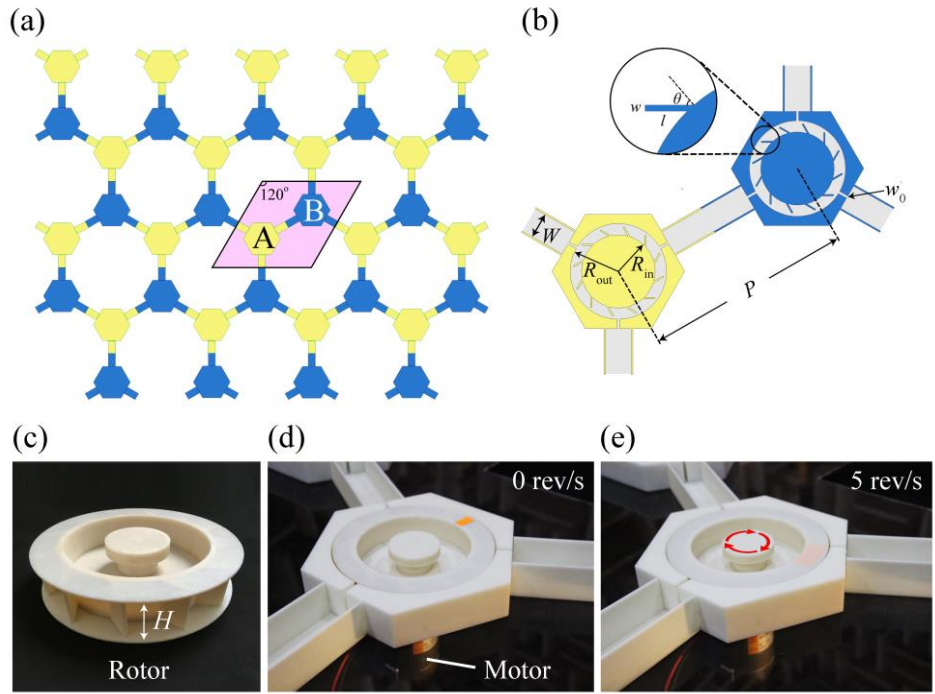

FIG. 1. (Color online) ACI with uniformly rotating ring resonators. (a) Schematic of the ACI with a honeycomb lattice. The primitive cell is shown by a shaded rhombus region, which contains two artificial 'atoms' (ring resonators) A and B. (b) Top view of a unit cell of the acoustic system. (c) Photograph of the chiral-structured rotor sample. (d-e) Photographs of motionless and rotating three-port ring resonators, respectively. The frequency of rotation is 5 rev/s.

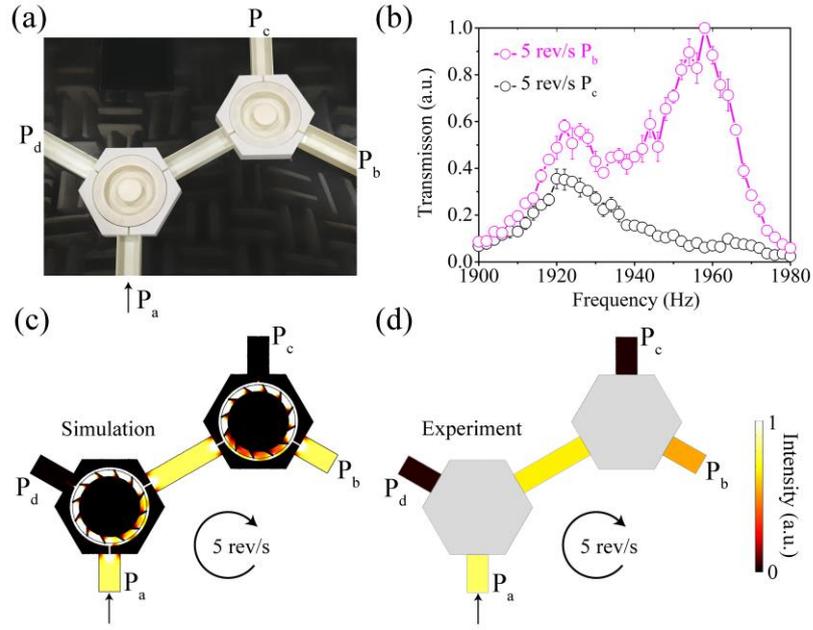

FIG. 2. (Color online) Nonreciprocal sound propagation in a unit cell. (a) The photograph of a unit cell of the acoustic system with four ports defined as $P_a$, $P_b$, $P_c$, and $P_d$, respectively. (b) The measured transmission spectra at ports $P_b$ and $P_c$. Acoustic waves are launched at port $P_a$ and the frequency of rotation is 5 rev/s. (c-d) The simulated and measured intensity distributions in the unit cell. The frequency of airborne sound is 1958 Hz.

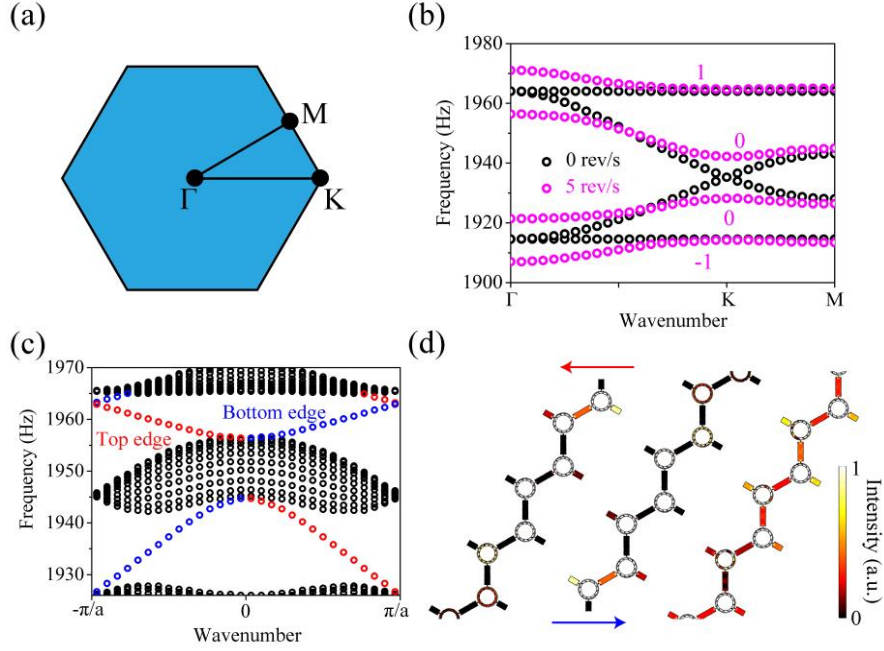

FIG. 3. (Color online) Bulk band structure and topological edge modes. (a) The first Brillouin zone of the honeycomb lattice. The irreducible Brillouin zone is marked by the triangle ΓMK. (b) Bulk band structure of ACI. The black and purple circles represent the band structures of unbiased and biased lattices, respectively, where the two-fold degeneracies at the Γ and K points for the motionless lattice are gapped on the condition of 5 rev/s rotation. For the biased lattice, the calculated Chern numbers of bulk bands are marked. (c) Band structure for a stripe super-cell of 10 biased resonators (5 rev/s). Black circles represent the bands of bulk modes, while blue and red circles correspond to the gapless edge modes with positive and negative group velocities, respectively. (d) Simulated intensity distributions of the edge modes (1958 Hz) and bulk modes (1967 Hz). The arrows indicate the propagation direction of one-way edge modes.

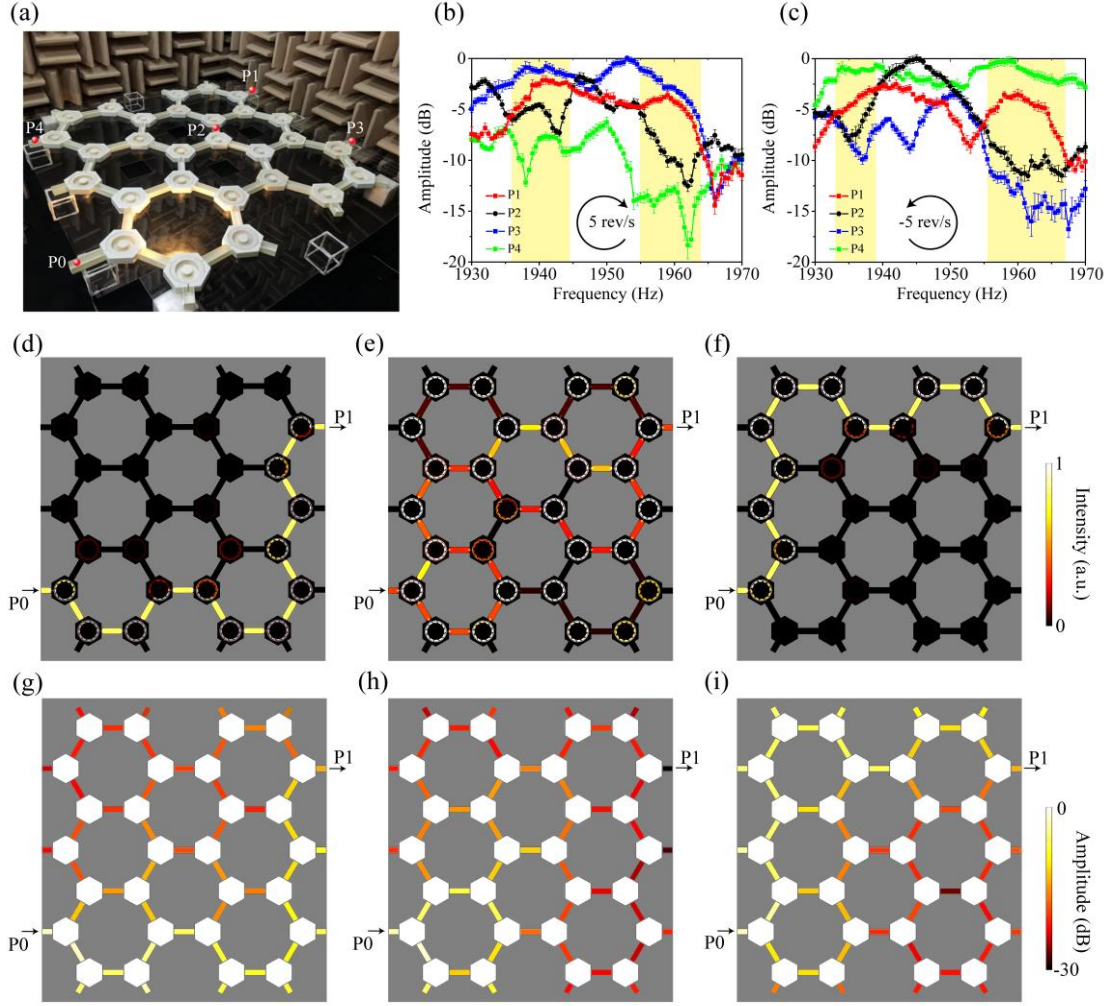

FIG. 4. (Color online) Experimental demonstration of ACI. (a) The photograph of ACI. The fabricated sample comprises 28 biased resonators and the size is 2.5 m×3 m. Ports P0, P1, P2, P3, and P4 are marked by the red dots. (b-c) Measured pressure amplitudes at ports P1, P2, P3, and P4 with the frequencies of rotation being 5 rev/s and −5 rev/s, respectively. The topological edge mode bands are highlighted by the yellow ribbons (1935.5~1944.5 Hz and 1955~1964 Hz in b; 1933~1939 Hz and 1955~1967.5 Hz in c). (d-f) Simulated intensity fields of anticlockwise chiral edge mode (1958 Hz, 5 rev/s), bulk mode (1967 Hz, 5 rev/s), clockwise chiral edge mode (1958 Hz, −5 rev/s), respectively. The acoustic waves are input at port P0 and output at port P1. (g-i) Measured pressure amplitude distributions of anticlockwise chiral edge mode (1958 Hz,

5 rev/s), bulk mode (1967 Hz, 5 rev/s), clockwise chiral edge mode (1958 Hz, −5 rev/s), respectively.